\documentclass[reprint, superscriptaddress]{revtex4-1}

\usepackage{graphicx, amsmath, amssymb, textcomp, siunitx, bm}
\begin{document}

\title{A digital twin for a chiral sensing platform}

\date{\today}

\author{Markus Nyman*}
\affiliation{Institute of Nanotechnology, Karlsruhe Institute of Technology}
\email{markus.nyman@kit.edu, carsten.rockstuhl@kit.edu}

\author{Xavier Garcia-Santiago}
\affiliation{Institute of Nanotechnology, Karlsruhe Institute of Technology}
\author{Marjan Krsti\'c}
\affiliation{Institute of Theoretical Solid-State Physics, Karlsruhe Institute of Technology}

\author{Lukas Materne}
\affiliation{Institute of Theoretical Solid-State Physics, Karlsruhe Institute of Technology}

\author{Ivan Fernandez-Corbaton}
\affiliation{Institute of Nanotechnology, Karlsruhe Institute of Technology}

\author{Christof Holzer}
\affiliation{Institute of Theoretical Solid-State Physics, Karlsruhe Institute of Technology}

\author{Philip Scott}
\affiliation{Institute of Applied Physics, Karlsruhe Institute of Technology}

\author{Martin Wegener}
\affiliation{Institute of Nanotechnology, Karlsruhe Institute of Technology}
\affiliation{Institute of Applied Physics, Karlsruhe Institute of Technology}

\author{Willem Klopper}
\affiliation{Institute of Physical Chemistry, Karlsruhe Institute of Technology}

\author{Carsten Rockstuhl*}
\affiliation{Institute of Nanotechnology, Karlsruhe Institute of Technology}
\affiliation{Institute of Theoretical Solid-State Physics, Karlsruhe Institute of Technology}
%\email{carsten.rockstuhl@kit.edu}

\begin{abstract}

Nanophotonic concepts can improve many measurement techniques by enhancing and tailoring the light-matter interaction. However, the optical response of devices that implement such techniques can be intricate, depending on the sample under investigation. That combination of a promise and a challenge makes nanophotonics a ripe field for applying the concept of a digital twin: a digital representation of an entire real-world device. In this work, we detail the concept of a digital twin with the example of a nanophotonically-enhanced chiral sensing platform. In that platform, helicity-preserving cavities with diffractive mirrors enhance the light-matter interaction between chiral molecules and circularly polarized light, allowing a faster measurement of the circular dichroism of the molecules. However, the sheer presence of the molecules affects the cavity's functionality, demanding a holistic treatment to understand the device's performance. In our digital twin, optical and quantum chemistry simulations are fused to provide a comprehensive description of the device with the molecules across all length scales and predict the circular dichroism spectrum of the device containing molecules to be sensed. Performing simulations in lockstep with the experiment will allow a clear interpretation of the results of complex measurements. We also demonstrate how to design a cavity-enhanced circular dichroism spectrometer by utilizing our digital twin. The digital twin can be used to guide experiments and analyze results, and its underlying concept can be translated to many other optical experiments.

\end{abstract}

\maketitle

\section{Introduction}

In standard spectroscopic experiments, the experimental conditions are usually simplified as much as possible to exclude sources of spurious signals, minimize measurement artifacts, and make the measurement repeatable and occur in the same conditions for each sample. For instance, when measuring the transmission spectrum of a sample of molecules in solution, the sample is held in a glass cell through which the light passes once. Quantities of interest, such as the absorption coefficient, are then calculated using appropriate formulae, such as Beer's law \cite{mayerhoferEmployingTheoriesFar2016}. Moving beyond such standard approaches, researchers have introduced various nanophotonic devices that enhance light-matter interaction to make more sensitive and/or selective measurements; many reviews have recently been published on this topic \cite{shakoorPortableNanophotonicSensors2019, zhangMetasurfacesBiomedicalApplications2021, el-khouryNanophotonicsChemicalImaging2022}. This progress towards nanophotonics-driven measurement technology is accompanied by challenges. Among these challenges, potentially the most important is that the interaction between the sample and the nanophotonic measurement device can be complex and often changes, for example, the shape of the spectrum being measured. In this situation, understanding the device-sample interaction as precisely as possible is vital and usually requires extensive computer simulations. Such needs prompt the development of what is nowadays called a digital twin.

Recently, the concept of digital twins has spread from industrial applications to science. Broadly speaking, a digital twin is a virtual replication of a real-world physical system, often combining simulation, data acquisition, and control software tools in a holistic framework \cite{lengDigitalTwinsbasedSmart2021}. A digital twin can be used to control and predict the performance of the real-world system based on data obtained from the system. In photonics and related fields, the concept has received attention in material processing \cite{zouDigitalTwinApproach2020, lunevDigitalTwinLaser2022}, imaging \cite{zhengFringeProjectionProfilometry2020}, design of light-emitting diodes \cite{vanderschansDigitalLuminaireDesign2020, ibrahimMachineLearningDigital2020}, and design of optical measurement setups \cite{vlaeyenDigitalTwinOptical2021}.

In this work, we address the problems of nanophotonics-enhanced spectroscopy experiments by introducing a digital twin that models the entire nanophotonic system and can act as an integral part of the experiment. We concentrate on a particular case, the sensing of chiral molecules through the measurement of their circular dichroism (CD) spectrum. The molecular CD is usually weak, both in absolute terms and when compared to the polarization-averaged absorption, and this traditionally results in long measurements times and the inability to measure minute concentrations. This is especially true for vibrational circular dichroism (VCD) on which we concentrate here \cite{mertenRecentAdvancesApplication2020, hermannQuantumCascadeLaserBased2022}. Hence, nanophotonic enhancement of the CD signal is an important research direction with potential for immediate applications \cite{luoPlasmonicChiralNanostructures2017, hentschelChiralPlasmonics2017, mohammadiNanophotonicPlatformsEnhanced2018, mohammadiAccessibleSuperchiralNearFields2019, solomonEnantiospecificOpticalEnhancement2019, garcia-guiradoEnhancedChiralSensing2020, warningNanophotonicApproachesChirality2021, bothNanophotonicChiralSensing2022, qinMetasurfaceMicroNanoOptical2022}. 

In our work, light-matter interaction enhancement is provided by helicity-preserving cavities. These are planar optical cavities defined by nanostructured mirrors that sustain cavity modes that approximately preserve optical helicity \cite{feisHelicityPreservingOpticalCavity2020, scottEnhancedSensingChiral2020, mauroChiralDiscriminationHelicitypreserving2023b}. A helicity-preserving cavity not only enhances the CD signal by a large factor but also changes the shape of the spectrum, often even flipping the sign of the actual CD signal, as shown later. This renders it a prime challenge to discriminate between the left- and right-handed enantiomers of a chiral molecule. A digital twin of this experiment solves this issue by providing a framework that considers all aspects of the experiment across all length scales in full detail. It constitutes a virtual environment that seamlessly supplements all measurements. Finally, it unlocks the opportunity to map between the sample at stake and the measured quantities, permitting an unambiguous interpretation. Beyond the methods behind the digital twin, we also demonstrate its applicability in a more complex setting, where many cavities are tuned to different operational wavelengths, and the task is to reconstruct the original CD spectrum of the bare molecule out of the individual measurement signals using machine-learning techniques. Such a spectral reconstruction is a task also encountered in many other types of nanophotonic spectrometry \cite{cerjanNanophotonicNoseCompressive2019, tittlImagingbasedMolecularBarcoding2018}. Ultimately, we believe this synthesis of experiment and computations allows us to utilize nanophotonic enhancement in precise measurements of physical quantities.

The article is structured as follows: in Section~\ref{sec-digital-twin}, we first discuss the preliminaries of CD spectroscopy and briefly recall the working principle of the helicity-preserving cavity with chiral molecules inside. We then describe the simulation methods that are seamlessly combined in the digital twin. In Section~\ref{sec-binol}, we apply the digital twin to a particular example of chiral sensing (binaphthol molecules) and elaborate on the parameters of the calculation. In Section~\ref{sec-spectrometer}, we take on the more complex example of reconstructing the entire spectrum of binaphthol from multiple cavity-enhanced signals. Section~\ref{sec-conclusions} presents concluding remarks.

\section{Digital twin for cavity-enhanced chiral sensing}
\label{sec-digital-twin}

\begin{figure*}
	\includegraphics[width=\linewidth]{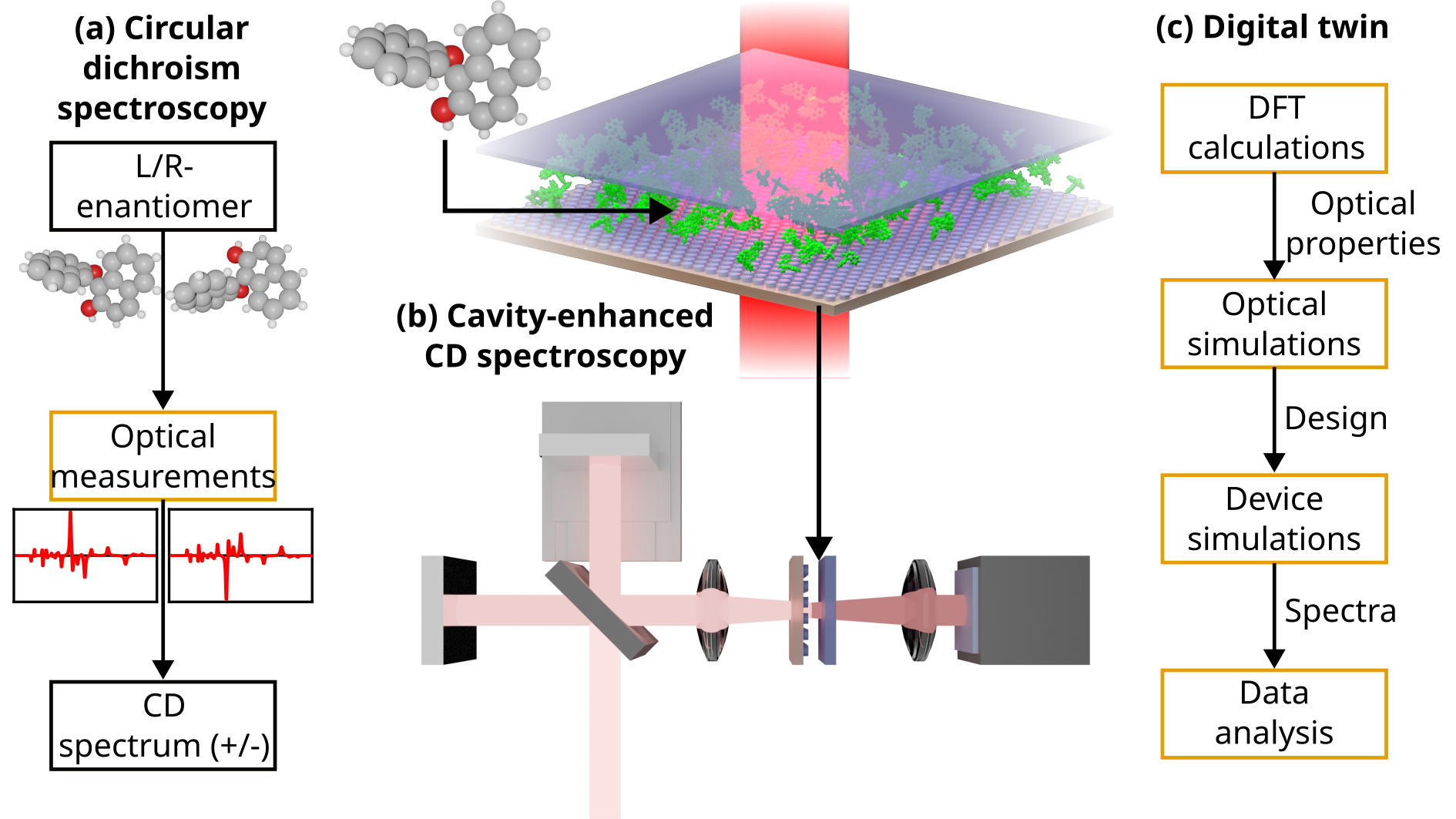}
	\caption{(a) Measurements of the circular dichroism (CD) of a chiral molecule can be used to distinguish between the molecule's two enantiomers because their CD spectra will have opposite signs. (b) In this work, cavity-enhanced CD measurements are performed by inserting the chiral molecules into a helicity-preserving optical cavity that enhances the light-matter interaction. This cavity can then be used instead of a normal sample cuvette in a CD spectrometer. The cavity enhancement modifies the shape of the spectrum, necessitating the development of a digital twin of the cavity-enhanced chiral sensing experiment to predict the results. (c) The digital twin replicates the experiment through a computer simulation and consists of four linked simulations. Density functional theory (DFT) simulations give the optical properties of the molecules. These are then used in optical simulations to design the cavity (or many cavities) to match the molecule. After simulating the response of the device for all wavelengths of interest, reference spectra are obtained, which can then be used in the data analysis phase.}
	\label{fig-overview}
\end{figure*}

The infrared CD spectrum of a molecule is typically measured using a Fourier transform infrared (FTIR) spectrometer. The sample is illuminated by a broadband beam of light whose polarization state is continuously modulated between left- and right-handed circular polarization and the transmission spectrum is then measured (in this work, we exclusively deal with transmission CD). The resulting signal can be separated into a constant part, proportional to the polarization-averaged transmittance of the sample, and an oscillating part, proportional to the CD signal \cite{lombardiObservationCalculationVibrational2009}. Let $T_\pm = P_\pm / P_0$ be the transmittance of left-handed (positive helicity) and right-handed (negative helicity) circular polarization; $P_\pm$ is the transmitted power and $P_0$ is the incident power. We define the CD signal as
\begin{equation}
\Delta T = T_{+} - T_{-} .
\end{equation}
As shown in Figure~\ref{fig-overview}(a), the left- and right-handed enantiomer of a molecule have CD spectra that are of the same shape, but the sign of the CD is flipped. If the sample has absorption coefficients $\alpha_\pm$ for the two polarizations, then
\begin{equation}
\Delta T = \exp(-\alpha_- d) - \exp(-\alpha_+ d) \approx \Delta \alpha d \exp(-\alpha_\text{avg} d) ,
\end{equation}
where $\Delta \alpha = \alpha_- - \alpha_+$ and $\alpha_\text{avg} = (\alpha_- + \alpha_+)/2$, and the approximation in the above equation is valid for $\Delta \alpha d << 1$. Here we have also ignored the reflections from the two surfaces of the sample. If the sample size is not limited, one can get the largest $\Delta T$ by setting $d = 1 / \alpha_\text{avg}$. For fairly high-concentration samples, this is practical. However, we aim to lower detection limits in this work, and thus, we will assume low concentrations. In such a setting, the optimal sample thickness $d$ is impractically large. Thus, we assume $d$ to be fairly small, in which case we have $\Delta T \approx \Delta \alpha d$.

Let us now consider a cavity-enhanced CD spectrometry experiment, illustrated in Figure~\ref{fig-overview}(b). The sample, molecules in solution, is inserted into a planar optical cavity that has nanostructured mirrors. We will describe this cavity, called a helicity-preserving cavity, in a later subsection. The cavity is inserted into the spectrometer in place of the usual cuvette sample holder, and the transmittance of the cavity-held sample is then measured. Obviously, the transmission spectrum is no longer the spectrum of the molecules but the spectrum of the whole device, including the molecules. As a figure of merit, we define a CD enhancement factor as
\begin{equation}
F = \frac{\Delta T}{\Delta T_\text{ref}} ,
\end{equation}
where $\Delta T$ is the CD signal of the device and $\Delta T_\text{ref} = \Delta \alpha d$.

Ultimately, we wish to detect small concentrations of chiral molecules, determine which enantiomer is present in the sample, and reconstruct the original spectrum of the molecules. For this, we need to model the system with quantitative accuracy, accounting for the interaction of the sample (the molecules) with the cavity that enhances the light-matter interaction. The key idea of this paper is to build a virtual representation of the entire experiment. We call this representation a digital twin of the experiment because it replicates the real-world system through computer simulations. The structure of our digital twin is shown in Figure~\ref{fig-overview}(c). It begins with using density functional theory (DFT) calculations to determine the optical properties of the chiral molecules. Knowing the optical properties, we proceed with designing and optimizing the cavities, for instance, to match with some of the peaks of the CD spectrum of the molecule. Once the cavities have been designed, we calculate their CD spectra that correspond to the spectra that would be measured in the real-world experiment. At the end of the data analysis phase, these spectra are used as a basis for interpreting experimental results, whether for discriminating between the two enantiomers of the molecule or even measuring the entire CD spectrum while still benefiting from cavity enhancement. In the following subsections, we elaborate on the most important aspects of our digital twin.

\subsection{Helicity-preserving optical cavity}

The key component of the physical setup is a helicity-preserving optical cavity. The helicity-preserving cavity is a planar optical cavity that sustains modes that preserve the circular polarization state of the incident light. The particular structure we use in this work is depicted in Figure~\ref{fig-optical-norminc}(a). The first mirror of the cavity is a nanostructured one consisting of a hexagonal array of silicon cylinders on a substrate. The second mirror is a flat silicon slab. The sample to be measured, chiral molecules in solution, is inserted in between the mirrors. As shown in Figure~\ref{fig-optical-norminc}(a), normally-incident light interacts with the nanostructured mirror and is diffracted into highly oblique angles. The nanostructured mirror is designed to preserve the helicity of the incident field. Moreover, when a circularly-polarized wave reflects from a (dielectric) interface at a large incidence angle, the polarization is not changed much, unlike in a normal-incidence Fabry-Perot resonator, where the polarization state would change upon each reflection \cite{feisHelicityPreservingOpticalCavity2020}. As a consequence, the cavity modes will have the same helicity as the incident light. The actual occurrence of the cavity modes sensitively depends on the cavity length, which is chosen to ensure the appearance of resonances. The helicity-preserving cavity is moreover achiral, which implies that, in the absence of fabrication imperfections, the CD signal is zero unless a chiral molecular material is inserted into the cavity. Therefore, any chiral signal seen in the transmitted light is due to the sample. The chiral properties of dielectric disc arrays have been previously studied in detail \cite{mohammadiNanophotonicPlatformsEnhanced2018, mohammadiAccessibleSuperchiralNearFields2019, ollanikHighlySensitiveAffordable2019, hanifehHelicityMaximizationDiffraction2020, hanifehOptimallyChiralLight2020, hanifehHelicityMaximizationPlanar2020, garcia-guiradoEnhancedChiralSensing2020}, and the characteristics of this type of cavity have been previously studied for normal-incidence waves \cite{feisHelicityPreservingOpticalCavity2020, scottEnhancedSensingChiral2020, baranovCircularDichroismMode2020}. CD enhancement factors up to 2000 were demonstrated theoretically with sample concentration-dependent spectral shapes.

\subsection{Overview of simulation methods}

The starting point of the digital twin must be the faithful representation of the optical properties of the chiral molecules. For that purpose, we rely on density functional theory (DFT) calculations. In this work, we use a development version of the TURBOMOLE V7.8 software package \cite{Franzke.Holzer.ea:TURBOMOLE.0} for electronic structure calculations to first perform geometry optimization of the molecule by calculating the ground state electron density. Starting from the optimized ground state electron density of the molecule in solvent, the molecular vibrational transitions are obtained, and the complex polarizabilities (electric dipole, magnetic dipole, and electric-magnetic dipole coupling) are computed by broadening the transitions with a Lorentzian function with a phenomenological damping parameter, usually in the range of 5 -- 10 cm$^{-1}$ for the half-width half-maximum (HWHM) to match the experimentally obtained spectrograms. A new algorithm for the damped dynamic polarizability tensors in the infrared part of the spectrum has been implemented in the TURBOMOLE package for this study, allowing us to construct T-matrices (discussed below) and efficiently bridge the scales from molecules to the device levels. The implemented algorithm is based on the work of Warnke and Furche \cite{warnkeCircularDichroismElectronic2012} and Zanchi {\it et al.} \cite{zanchiEvaluationMolecularPolarizability2019}. It uses sum-over-states calculation of the complex frequency dependent polarizability tensors from atomic polar tensors (APTs) and atomic axial tensors (AATs) transformed to the normal vibration modes of the molecule. The electric-electric, electric-magnetic, and magnetic-magnetic polarizability tensors can be calculated as: \cite{warnkeCircularDichroismElectronic2012,materneDevelopmentApplicationComputational2022}
\begin{align}
\label{eq_alpha_ee}
& \bm{\alpha}^{e,e}(-\omega, \omega) = & \frac{1}{\hbar} \sum_a \frac{\langle 0_a | \bm{\mu} | 1_a \rangle \otimes \langle 1_a | \bm{\mu} | 0_a \rangle}{\omega_a - \omega - i \Gamma / 2} \\
& & + \frac{1}{\hbar} \sum_a \frac{\langle 0_a | \bm{\mu} | 1_a \rangle \otimes \langle 1_a | \bm{\mu} | 0_a \rangle}{\omega_a + \omega - i \Gamma / 2} , \\
\label{eq_alpha_em}
& \bm{\alpha}^{e,m}(-\omega, \omega) = & \frac{1}{\hbar} \sum_a \frac{1}{\omega_a} \frac{\langle 0_a | \bm{\mu} | 1_a \rangle \otimes \langle 1_a | \bm{m} | 0_a \rangle}{\omega_a - \omega - i \Gamma / 2}\\
& & - \frac{1}{\hbar} \sum_a \frac{1}{\omega_a} \frac{\langle 0_a | \bm{\mu} | 1_a \rangle \otimes \langle 1_a | \bm{m} | 0_a \rangle}{\omega_a + \omega - i \Gamma / 2} , \\
\label{eq_alpha_mm}
& \bm{\alpha}^{m,m}(-\omega, \omega) = & \frac{1}{\hbar} \sum_a \frac{\langle 0_a | \bm{m} | 1_a \rangle \otimes \langle 1_a | \bm{m} | 0_a \rangle}{\omega_a - \omega - i \Gamma / 2}\\
& & + \frac{1}{\hbar} \sum_a \frac{\langle 0_a | \bm{m} | 1_a \rangle \otimes \langle 1_a | \bm{m} | 0_a \rangle}{\omega_a + \omega - i \Gamma / 2} ,
\end{align}
where $\omega$ is the frequency of interest, $\omega_a$ is the frequency of the a$^\mathrm{th}$ mode, $\Gamma$ is the damping parameter with $\Gamma / 2$ being equivalent to the half width at half maximum (HWHM), $\otimes$ is the outer product operator, and\cite{nicuVibrationalCircularDichroism2008,Reiter.Kuehn.ea:Vibrational-circular-dichroism.2017}
\begin{align}
& \langle 0_a | \bm{\mu} | 1_a \rangle = \langle 1_a | \bm{\mu} | 0_a \rangle = \left(\frac{\hbar}{\omega_a}\right)^{1/2} \sum_j \frac{\partial \bm{\mu}}{\partial x_j} L_{ja} \\
& \langle 0_a | \bm{m} | 1_a \rangle = -\langle 1_a | \bm{m} | 0_a \rangle = - \left(2 \hbar \omega_a \right)^{1/2} \sum_j \frac{\partial \bm{m}}{\partial x_j} L_{ja}.
\end{align}
Here, $\mathbf{L}$ is a transformation matrix that relates the Cartesian displacements $x_i$ of the atoms to the normal coordinates. The operators $\partial \bm{\mu} / \partial x_i$ encompass the APTs, while $\partial \bm{m} / \partial x_i$ gather the AATs \cite{stephensTheoryVibrationalCircular1985, stephensGaugeDependenceVibrational1987, nicuVibrationalCircularDichroism2008}. Eqs.~\ref{eq_alpha_ee} to \ref{eq_alpha_mm} have been implemented in the course of this work. Contrary to electronic excitation, where these equations are tedious to evaluate directly, the limited number of vibrationally excited states makes the sum-over-states approach feasible. It is therefore straightforward to implemented eqs.~\ref{eq_alpha_ee} to \ref{eq_alpha_mm} in any suitable program that can evaluate the corresponding vibrational APT and AAT tensors.  Note that the magnetic-magnetic part, $\bm{\alpha}^{m,m}$, is listed and implemented for completeness, but neglected in this work due to its vanishingly small contribution to the T-matrices.

We conduct both the ``optical simulations'' and ``device simulations'' steps shown in Figure~\ref{fig-overview}(c) using the T-matrix method. The T-matrix describes the linear light-matter interaction of a single scatterer, connecting the incident and scattered fields, both of which are represented using vector-spherical-harmonic fields. The T-matrix of a molecule can be obtained from polarizability tensors calculated on the precise quantum level of theory by the DFT-based method \cite{fernandez-corbatonComputationElectromagneticProperties2020}, while the T-matrix of a classical scatterer, such as one of the cylinders in a helicity-preserving cavity, is calculated using a Maxwell's equations solver \cite{beutelEfficientSimulationBiperiodic2021}. For these computations, we utilize the finite element simulation package JCMsuite \cite{JCMSuite2023}. The T-matrix serves as the starting point for calculating the plane-wave transfer matrix of each of the components of the system, and the optical response of the whole system can be obtained by stacking the transfer matrices together \cite{beutelEfficientSimulationBiperiodic2021, zerullaMultiScaleApproachModeling2022}. These calculations are performed using the \texttt{treams} Python package \cite{Treams2023}. The main advantage of T-matrix-based simulations is speed: once the T-matrices of all parts of the system have been calculated, the response of the whole system can be computed quickly. Also, parameters such as the period of the cylinder array and the length of the cavity can be varied without calculating new T-matrices. This permits the large parameter sweeps of the following section.

\section{Sensing BINOL molecules}
\label{sec-binol}

\begin{figure*}
	\includegraphics[width=\linewidth]{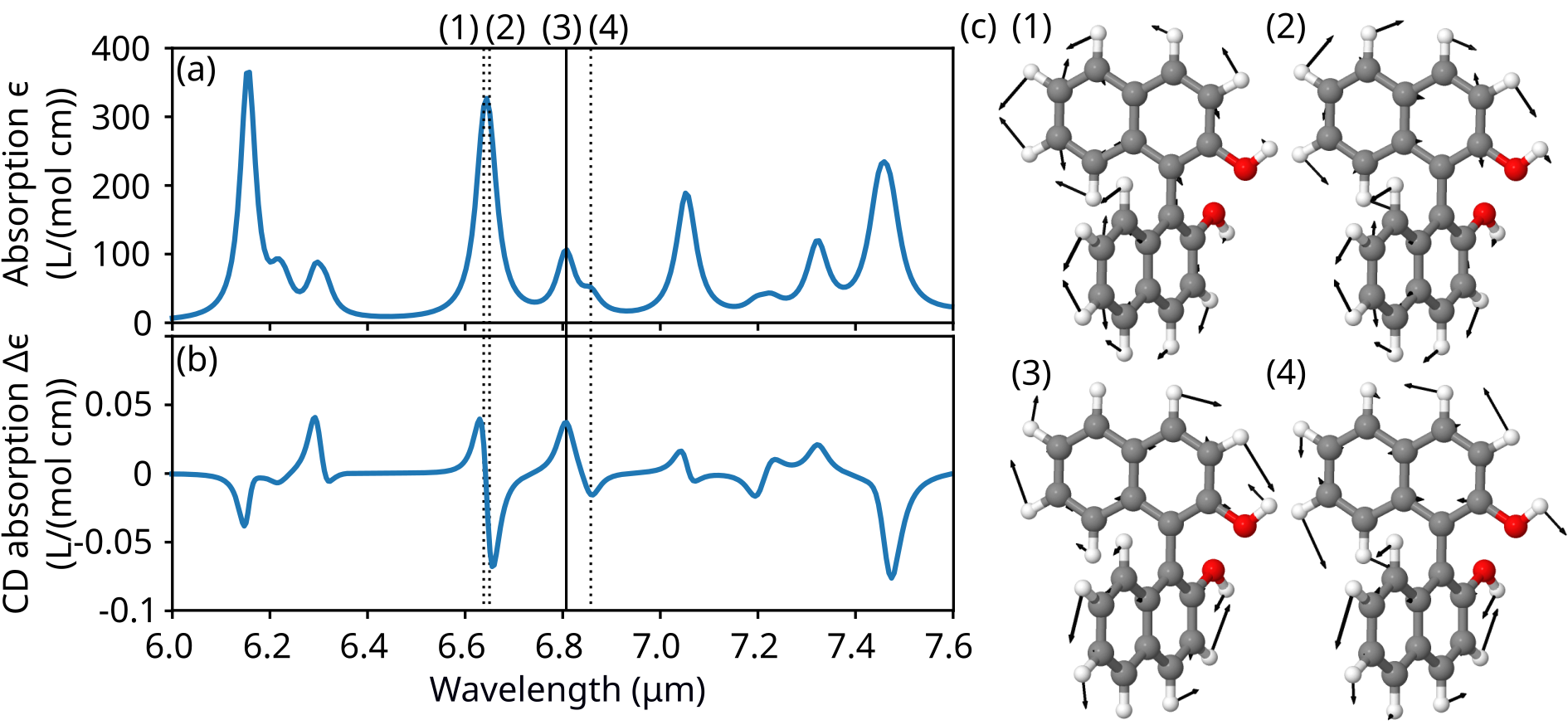}
	\caption{The optical properties of chiral molecules are calculated from the T-matrices obtained from the density functional theory approach. Panels (a) and (b) show, respectively, the calculated molar absorption coeeficients for the infrared ($\epsilon$) and VCD ($\Delta \epsilon$) spectrum of the BINOL molecule in chloroform solvent, which were used as a benchmark example. A vertical solid line marks the chiral resonance at \SI{6.805}{\micro\meter} wavelength. The dotted lines also highlight three surrounding resonances. The displacement vectors of the selected molecular vibrations that these resonances correspond to are shown in (c).}
\label{fig-binol}
\end{figure*}

We now illustrate the function and capabilities of the digital twin by taking a particular example, the sensing of 1,1'-Bi-2-naphthol (BINOL) molecules. The BINOL molecule was selected because it is widely available and extensively used in the context of electronic and vibrational CD spectroscopy due to its strong chiral properties. In the following, we will use the digital twin to simulate the optical response of a realistic concentration of BINOL molecules, design a helicity-preserving cavity to match one of the molecule's resonances and calculate the CD spectrum.

We begin by conducting DFT calculations on the (R)-(+)-BINOL enantiomer, assuming the molecules are dissolved in chloroform. The details of the DFT calculations can be found in the Supplementary Material. Briefly, the geometrical shape of the molecule is optimized by a gradient descent-based method, and surrounding solvent effects were considered by an implicit conductor-like screening model (COSMO) as implemented in TURBOMOLE for chloroform. A stable structural minimum of the BINOL molecule was confirmed by analysis of the vibrational modes, followed by the calculation of the polarizability tensors in the middle infrared, the most informative part of the vibrational spectrum. The constructed T-matrices from polarizabilities were used to simulate the absorption and CD spectra as depicted in Figs.~\ref{fig-binol}(a) and (b), respectively. The overall spectral shape matches well to the experimentally recorded spectra reported previously \cite{nicuUnderstandingSolventEffects2012}. The chosen damping of 10 cm$^{-1}$ for the HWHM of the Lorentzian lineshape convolves the discrete transitions and simulates the thermally-induced broadening of the spectrum in the experiment. The absorption spectrum exhibits four intensive peaks at 6.15, 6.65, 7.05, and \SI{7.48}{\micro\meter}, with many lower intensity transitions in between. The most intensive CD signal is for transitions at 6.65, 6.805, and \SI{7.48}{\micro\meter}. Those vibrational transitions exclusively correspond to modes where single and double carbon-carbon bonds in aromatic rings and hydrogen-saturated C-O bonds are stretching and bending. In Figure~\ref{fig-binol}(c), the displacement vectors of atoms of four selected vibrational modes corresponding to the interesting modes are visualized by black arrows superimposed on the figure of the molecular structure. Displacement vectors of hydrogen atoms manifesting symmetric and asymmetric rocking on the carbon and oxygen atoms accompanying C-C stretching and bending in the molecules have larger displacement vectors because they are much lighter than the C and O atoms. Overall, spectral features of both IR and CD spectroscopy give us the confidence to proceed further with the design of the enhancing cavity for the CD transition at \SI{6.805}{\micro\meter}. That peak was selected for enhancement due to the reasonably high ratio between CD and normal absorption. This can be beneficial for high-concentration samples, as shown in \cite{scottEnhancedSensingChiral2020}.

\begin{figure*}
	\includegraphics[width=\linewidth]{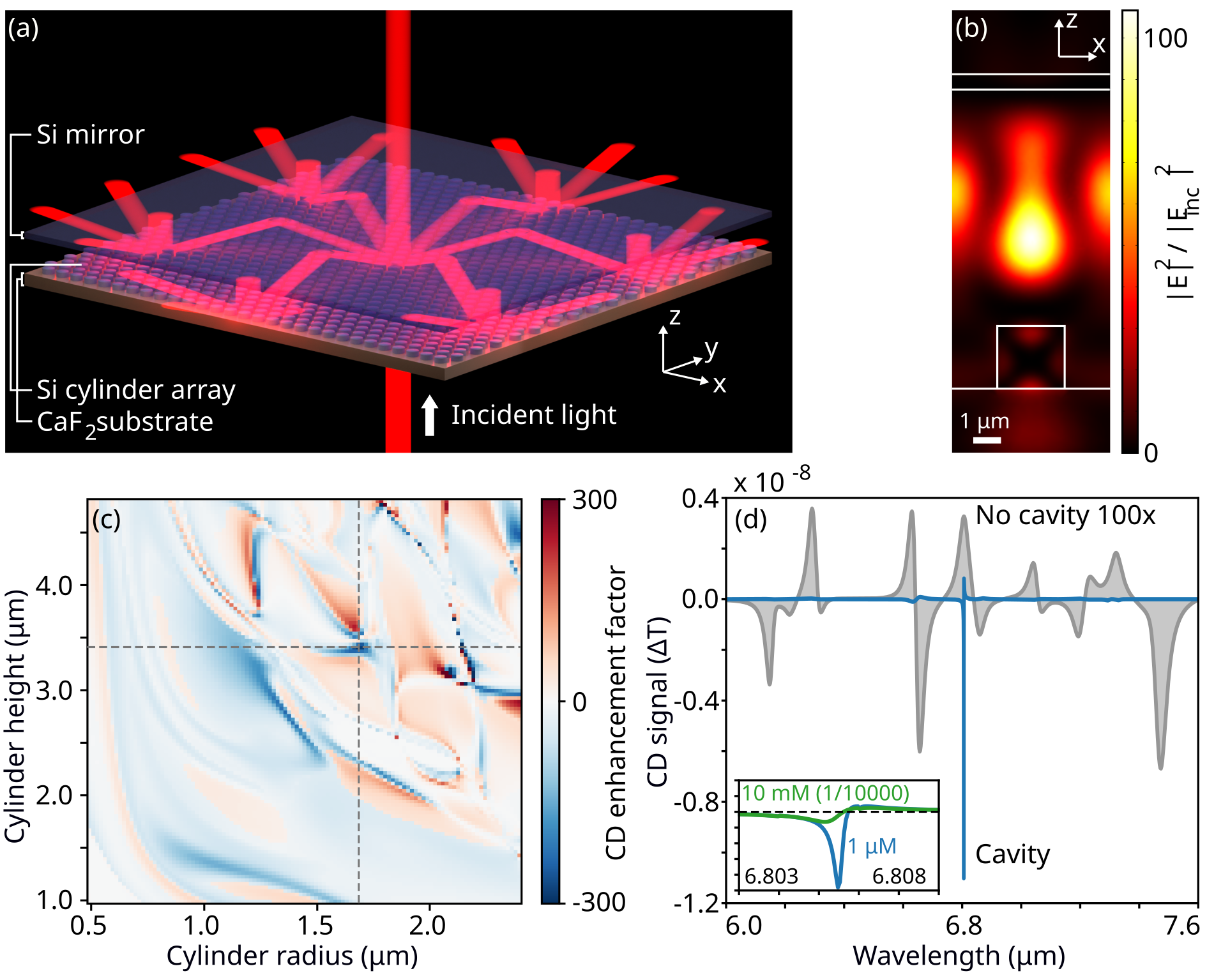}
	\caption{A helicity-preserving cavity is illustrated in (a). Its structure consists of two mirrors lying on two separate calcium fluoride substrates (the upper substrate is omitted from the figure). The bottom mirror consists of a hexagonal lattice of silicon cylinders, while the top mirror is a simple thin film of silicon. The solution containing the molecules to be analyzed is inserted in between the mirrors. The red beams of light illustrate how the cavity functions: light enters the cavity through the silicon disc array, which splits the beam into diffraction orders that couple into the cavity modes. Panel (b) shows a cross-section of the cavity together with the electric-field intensity distribution for a normally-incident circularly polarized plane wave. A strong intensity enhancement is obtained, and the interference of the diffraction modes is visible as a standing wave that modulates the intensity distribution along the $x$-direction. To optimize the cavity for the given \SI{6.8}{\micro\meter} design wavelength, we explore the parameter space consisting of cylinder radius $R$ and height $H$ together with the cavity length $L$. Panel (c) shows a pseudocolor plot of the circular dichroism enhancement factor of the cavity for a range of $R$ and $H$. In each pixel in the plot, the optimal cavity length $L$ is chosen, and the corresponding CD enhancement value is plotted. There are many local optima with an enhancement factor of about 300, and the one at $R = $ \SI{1.7}{\micro\meter} and $H = $ \SI{3.4}{\micro\meter} is chosen. Panel (d) shows the CD spectrum of this optimal cavity (blue curve) compared to the CD spectrum of the molecules alone (gray curve); the latter is multiplied by a factor of 100 to bring it to the same scale with the enhanced spectrum. At the design wavelength, a 300-fold enhancement is obtained. The inset shows a narrow part of the spectrum around the resonance. To illustrate the dependence of the system's response on the sample, the green curve shows the CD spectrum obtained from the optimal cavity if the concentration of BINOL molecules is increased to \SI{10}{\milli\mole/\liter}. The lineshape of the resonance is then flattened.}
	\label{fig-optical-norminc}
\end{figure*}

To illustrate the field distribution inside the cavity at the chosen wavelength, Figure~\ref{fig-optical-norminc}(b) shows the light intensity in an optimized cavity filled only with solvent. The excitation is with a normally-incident plane wave with left-handed circular polarization, and the cross-section in the $xz$-plane is shown. As expected, we observe the interference of the diffraction orders as a standing wave along the $x$-direction and the interference of up- and down-propagating waves as a standing wave in the $z$-direction. We also see a peak local intensity enhancement factor of 110.

To optimize this cavity, we performed optical simulations using the T-matrix-based method. We fixed the wavelength for the optimization to \SI{6.8}{\micro\meter}. We require the diffraction orders to propagate at $\theta_\text{d}$ = 80$^\circ$ angle, which then gives us the required period of the cylinder array from the diffraction equation for a hexagonal lattice $\Lambda = 2 \lambda_0 / (\sqrt{3} n \sin(\theta_\text{d}))$. We therefore set $\Lambda =$~\SI{5.684}{\micro\meter}. The choice of the diffraction angle presents a tradeoff between a higher quality factor (larger angle) and larger bandwidth and lower loss through diffracted light escaping from the side of the cavity (smaller angle). We also fix the thickness of the second mirror (the silicon slab) to \SI{0.55}{\micro\meter}. We then varied the remaining geometrical parameters: cylinder radius $R$, cylinder height $H$, and cavity length $L$, in each case calculating the transmission CD, $\Delta T$, of the cavity for a normal-incidence plane wave. This is easily done in a brute-force manner because full-wave simulations are only required to obtain the T-matrices of the cylinders. The T-matrix method is comparatively fast and energy efficient, allowing us to calculate vast parameter sweeps quickly. Figure~\ref{fig-optical-norminc}(c) shows the CD enhancement factor as a function of $R$ and $H$. The best $L$ for the corresponding ($R$, $H$) is chosen at each point on the color plot. Best enhancement factors are obtained for moderately large cylinders, and there are many local optima. We note, however, that many of these local optima are very narrow features in the plot. This would make the structure exceedingly sensitive to fabrication imperfections, and thus, we choose to use $R$ = \SI{1.7}{\micro\meter}, $H$ = \SI{3.4}{\micro\meter} with $L$ = \SI{8.7}{\micro\meter}. Here the feature is fairly resistant against small perturbations in $R$ and $H$. The corresponding CD enhancement reaches 300.

Figure~\ref{fig-optical-norminc}(d) shows the CD spectrum of the designed cavity with BINOL molecules inside at a concentration of \SI{1}{\micro\mole/\liter} (blue curve). The spectrum peaks sharply at the design wavelength. For comparison, the CD spectrum of the same sample without the cavity is shown in gray; this curve is multiplied by a factor of 100 to make it visible. This verifies the CD enhancement and also illustrates how drastically the shape of the spectrum changes. For example, the sign of the CD is even flipped by the cavity. Furthermore, the inset of this figure shows the lineshape for two different concentrations: \SI{1}{\micro\mole/\liter} (blue curve) and \SI{10}{\milli\mole/\liter}. The flattening of the lineshape for the higher concentration is expected because of increased absorption, which reduces the Q-factor of the resonance. It also illustrates that, in general, the output signal cannot be expressed as a product of the plain molecule's CD spectrum with an enhancement factor spectrum. Therefore, simulating the system with the cavity and the sample together is necessary. In the workflow discussed in Section~\ref{sec-digital-twin}, this would mean performing a simulation for each measurement, making the physical and virtual experiments intimately connected.

\begin{figure*}
	\includegraphics[width=\linewidth]{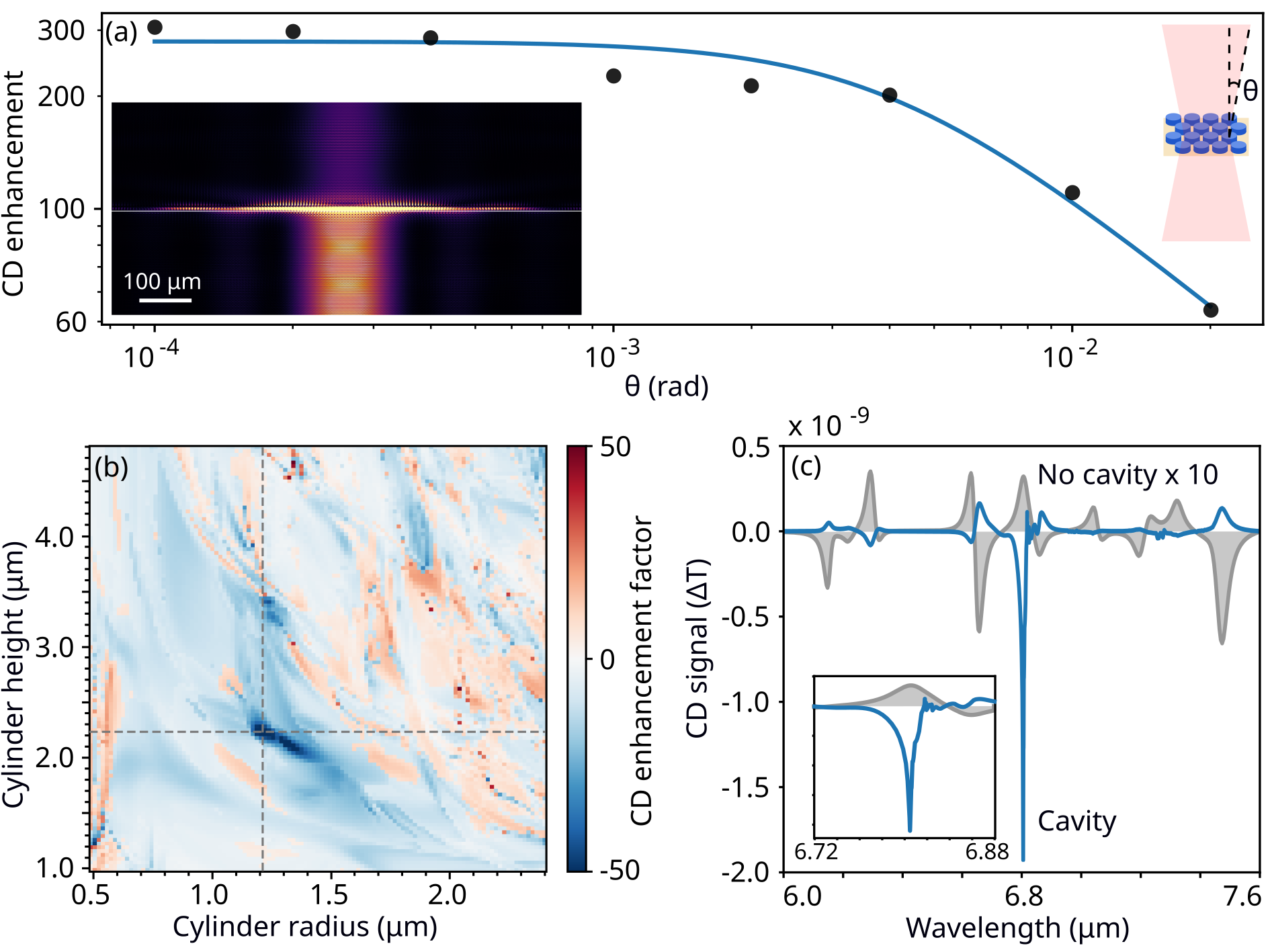}
	\caption{Optimization of CD enhancement while accounting for the finite size of the CD-enhancing cavity, a typical experimental limitation. The right inset of panel (a) illustrates that light passing through the cavity diverges, necessitating the consideration of an angular spectrum of plane waves with divergence angle up to $\theta$. The angular spread leads to smaller CD enhancement factors than those shown in Figure~\ref{fig-optical-norminc}, as shown in panel (a), which plots the maximum available CD enhancement against $\theta$. Beyond a critical value of $\theta$, the enhancement decreases proportional to $1/\theta$, and the enhancement no longer reaches the limit set by the cavity's finesse. The optimization of the cavity is done in the same way as in Fig.~\ref{fig-optical-norminc}, but here, we set $\theta = $ 0.02 rad to account for the divergence of light passing through a \SI{250}{\micro\meter}-wide cavity; the lower left inset of panel (a) shows the intensity distribution of the beam. Panel (b) shows the optimization surface with cylinder radii $R$ and heights $H$. In each pixel of the image, the optimal $L$ is chosen. An optimum is found at $R = $ \SI{1.2}{\micro\meter} and $H = $ \SI{2.25}{\micro\meter}, where the enhancement factor reaches -50 and is not too sensitive to the exact values of $R$ and $H$, making the cavity more robust against, e.g., fabrication imperfections. For this design, cavity length $L = $ \SI{8.58}{\micro\meter} is optimal. Panel (c) shows the circular dichroism spectra of a \SI{1}{\micro\mole/\liter} solution of BINOL inside the cavity (blue line) and outside the cavity for the same sample thickness (orange line). The latter is multiplied by a factor of 10 to bring it to the same scale as the enhanced spectrum. The signal from the cavity is 60 times larger than from the plain molecules. The inset shows the line shape in a narrower spectral region.}
	\label{fig-optical-divergence}
\end{figure*}

The simulations shown in Fig.~\ref{fig-optical-norminc} illustrate the ideal case, but are somewhat simplistic because just one plane wave is used as input. We now take into account the experimental limitations, in particular, the limited size of the cavity. This is done by considering the transmission of an optical beam with a finite divergence angle $\theta$, which results in a limited beam spot size. The actual edge effects are still ignored. For simplicity, we consider beams with a flat, constant-intensity angular spectrum. The effect of increasing the divergence angle $\theta$ on the peak CD enhancement factor is shown in Fig.~\ref{fig-optical-divergence}(a). For small $\theta$, enhancement factors of about 300 are obtained, just like in Fig.~\ref{fig-optical-norminc} for the normal-incidence case. Once a certain threshold $\theta$ is crossed, the CD enhancement decreases because many of the plane waves that constitute the optical beam are no longer in resonance with the cavity. The blue line shows a fit described by $a F_\text{max} / \sqrt{a^2 + \theta^2}$, with the values $a = 0.004$ rad and $F_\text{max} = 280$. In the limit of large $\theta$, the fit shows a $1/\theta$ behavior. This is consistent with the behavior of planar resonators for oblique incidence angles; we prove this in the Supplementary Information.

We perform the optimization as in Fig.~\ref{fig-optical-norminc}, but use a beam with $\theta = 0.02$ rad expanded into a 9-by-9 grid of plane waves. This multiplies the computational effort, but the calculations remain tractable owing to the speed of the T-matrix-based calculation method. The chosen width of the angular spectrum corresponds to a beam with a spot size of about \SI{250}{\micro\metre}. Therefore, we predict that a cavity with a lateral size on the order of a few hundred \textmu m would be sufficient for experimental verification of the CD enhancement presented here. Figure~\ref{fig-optical-divergence}(b) shows the CD enhancement factor for a range of $R$ and $H$. As previously, in each point in the plot, we pick the best $L$. As expected, the increased divergence reduces the CD enhancement, but we also obtain a different optimum: $R$ = \SI{1.2}{\micro\meter} and $H$ = \SI{2.25}{\micro\meter} with $L$ = \SI{8.55}{\micro\meter}. Here, we get an enhancement factor of 50, which is also very stable against small perturbations of $R$ and $H$. If we now insert a \SI{1}{\micro\mole/\liter} solution of BINOL into the cavity, we obtain the CD spectrum shown in Fig.~\ref{fig-optical-divergence}(c). As expected, it is broader than in the previous example, but still reaches a respectable CD enhancement of 50. We also see that the sign of CD is again changed.

Spectra such as those shown in Fig.~\ref{fig-optical-norminc}(d) and Fig.~\ref{fig-optical-divergence}(c) can be measured using a CD spectrometer. The digital twin can then be used to produce the corresponding calculated spectra to which the experimental ones may be compared. For the simple task of detection and enantiomer discrimination of a single type of molecule, using just one helicity-preserving cavity tuned to the strongest molecular resonance may be sufficient. Preferably, the light source should also match the target wavelength because the CD enhancement outside the target wavelength is fairly modest. However, using only one cavity would not allow one to distinguish between multiple different molecules efficiently. This issue can be solved by using multiple cavities. To demonstrate this, we proceed to our final example, which is easy to construct thanks to the digital twin: a cavity-enhanced CD spectrometer.

\section{Cavity-enhanced CD spectrometer}
\label{sec-spectrometer}

Due to the narrow linewidth of helicity-preserving cavities that provide a significant CD enhancement, it becomes necessary to create many cavities that span a larger range of wavelengths together. This allows one to distinguish between many different molecules and their enantiomers. We shall also show that, by applying a reconstruction algorithm, the original spectrum of a molecule may be retrieved by simply measuring the total transmitted power going through each cavity.

\begin{figure*}
	\includegraphics[width=\linewidth]{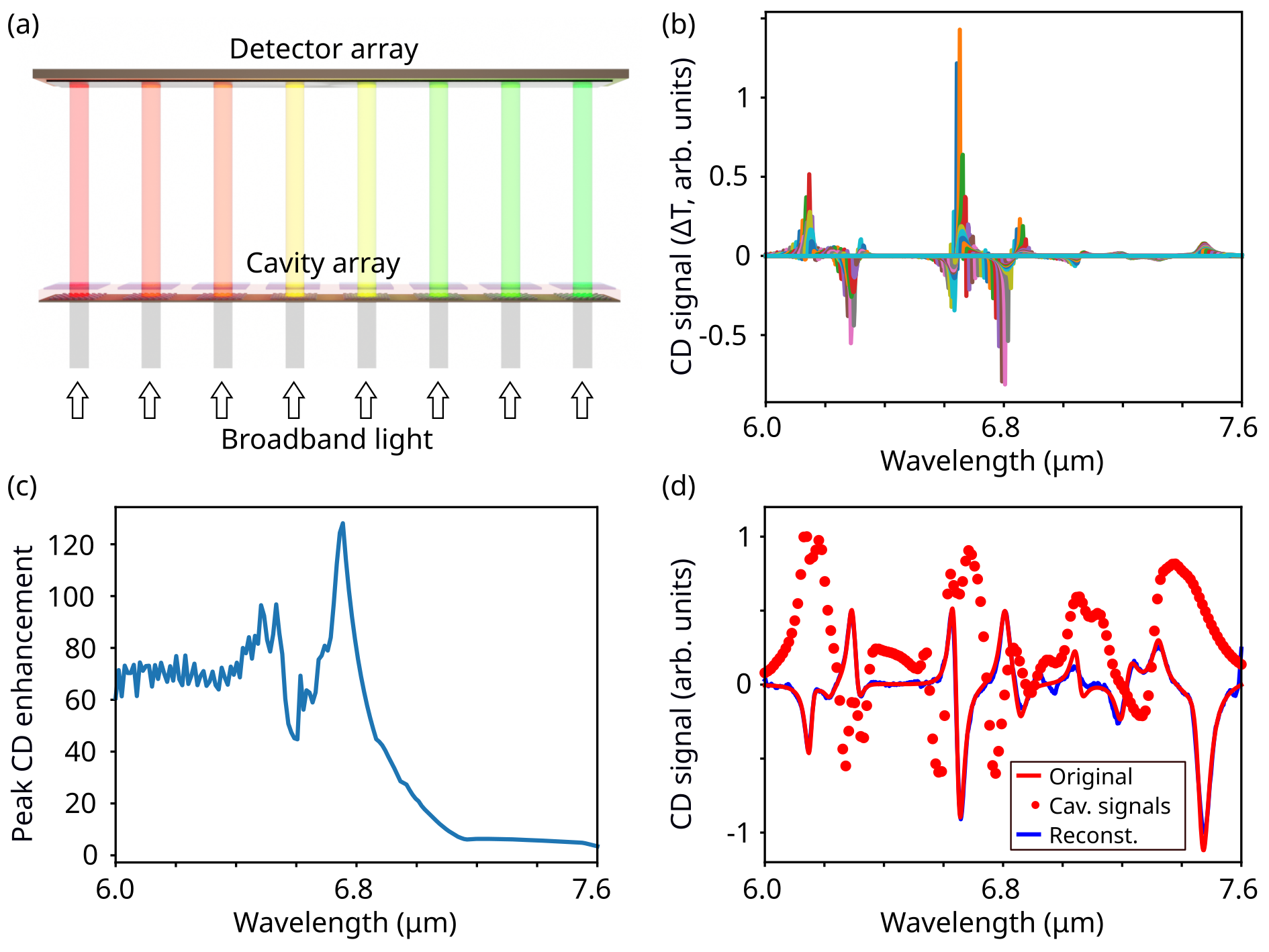}
	\caption{An array of helicity-preserving cavities tuned to different wavelengths can act as a spectrometer. Panel (a) shows an illustration of the array. We assume that the cavities would be fabricated on a single substrate in one run, and thus we regard the height of the cylinders $H$ and the cavity length $L$ to be fixed parameters while allowing the array period $\Lambda$ to vary from one cavity to the next. Panel (b) shows the CD spectrum of BINOL at \SI{1}{\micro\meter} concentration, stitched together from the spectra of 160 cavities, their center wavelengths spanning the wavelength range \SI{6.0}{\micro\meter} to \SI{7.6}{\micro\meter}. Panel (c) shows the maximum absolute CD enhancement factors of each cavity, demonstrating that enhancement factors over 60 are obtained for a wide range of wavelengths and the enhancement peaks at 125. Finally, panel (d) shows the reconstruction (blue solid line) of BINOL's circular dichroism spectrum (red solid line) from measurements of the total intensity transmitted by each cavity (red dots). The reconstructed spectrum is mostly identical to the original spectrum, showing that the cavity-enhanced spectrometer works well.}
	\label{fig-spectrometer}
\end{figure*}

Figure~\ref{fig-spectrometer}(a) illustrates the cavity-enhanced CD spectrometer. It consists of an array of helicity-preserving cavities, each is tuned to a different operational wavelength. As before, the liquid sample to be measured is inserted inside the cavities. The cavities are illuminated all together by broadband light, and on the transmission side, the optical power transmitted through each cavity is measured by an array detector such as a photodiode array. In this example, we used the digital twin to design 160 cavities spanning the wavelength range \SI{6.0}{\micro\meter} to \SI{7.6}{\micro\meter}. This was accomplished by taking the design used in the example of Fig.~\ref{fig-optical-divergence} and varying the cylinder array period $\Lambda$. We assume that the cylinder arrays would be fabricated in one run on a single substrate and, likewise, a single substrate with a large-area Si thin film would be used as the second mirror. In this case, the cylinder heights $H$ would be fixed by the amount of Si deposited, and the cavity length $L$ would be the same for all cavities. While varying $R$ would still be possible and would slightly improve the results shown here, it makes no difference from the point of view of a proof of concept.

Assuming each of the cavities is filled with \SI{1}{\micro\mole/\liter} of BINOL in chloroform, Fig.~\ref{fig-spectrometer}(b) shows the spectra of all the cavities together. In the measurement process where the total power is measured, each cavity-specific spectrum is integrated over all wavelengths. To complicate matters, the enhancement factor of the cavities is not the same. Figure~\ref{fig-spectrometer}(c) shows the peak CD enhancement factor as a function of the center wavelength of the cavity. Here we see that the enhancement ranges from 60 to 120 between 6.0 and \SI{6.8}{\micro\meter} and then drops to low values for longer wavelengths; this is mainly because the cavity length is no longer optimal for these wavelengths.

As a proof of principle, let us consider an optically thin sample for which it is possible to write the transmitted CD signal of the $n^{\mathrm{th}}$ cavity as
\begin{equation}
s_n = \int F_n(\lambda_0) \Delta T_\text{molec}(\lambda_0) d\lambda_0 ,
\end{equation}
where $F_n(\lambda_0)$ is the spectrum of the CD enhancement and $\Delta T_\text{molec}(\lambda_0)$ is the molecule's CD spectrum. If we now discretize the spectra using a constant wavelength step $\Delta \lambda$, this integral turns into a sum
\begin{equation}
s_n = \Delta \lambda \sum_i F_{n,i} u_{i}, \label{eq-cavitysignaldiscrete}
\end{equation}
where the index $i$ runs over the discrete wavelength and for the sake of notational simplicity $u_i = \Delta T_\text{molec}(\lambda_i)$. We now want to estimate $\Delta T_\text{molec}(\lambda)$. The system of $n$ equations from Equation~\ref{eq-cavitysignaldiscrete} can be expressed as a matrix equation $\mathbf{s} = \mathbf{F} \mathbf{u}$. The reconstruction requires us to find the inverse of $\mathbf{F}$, which, in general, does not exist as the matrix is not square. However, we can find an approximate inverse, the matrix $\mathbf{G}$, that minimizes the error $|\mathbf{u}_\text{rec} - \mathbf{u}| = |\mathbf{G} \mathbf{s} - \mathbf{u}|$ for physically-plausible spectra. To find the reconstruction matrix $\mathbf{G}$, we first calculate $\mathbf{F}$ using the digital twin. Then, we treat the system $\mathbf{u} = \mathbf{G} \mathbf{s}$ as a neural network with one fully-connected layer, no bias vector, and linear activation function. We generate random spectra $\mathbf{u}$ that consist of sums of Lorentzian functions, calculate the corresponding cavity signals $\mathbf{s}$ using Eq.~\eqref{eq-cavitysignaldiscrete}, and use this as training data to find the elements of $\mathbf{G}$ using the TensorFlow library. It is crucial to note that the spectrum of BINOL or any other molecular spectrum is not present in the training data. The training process is described in more detail in the Supplementary Information.

We now use the optimized $\mathbf{G}$ to retrieve the spectrum of BINOL. In Fig.~\ref{fig-spectrometer}(d), the original spectrum is the red solid line. The cavity signals are the red dots. The reconstructed spectrum is shown by the blue solid line. In this case, the reconstruction is very good, except for an edge artifact at \SI{7.6}{\micro\meter} and underestimation of the height of the peak near \SI{7.1}{\micro\meter}. This shows that even though the individual cavity enhancement spectra have many features beyond the main spike as seen in Fig.~\ref{fig-optical-divergence}(c), the reconstruction is not adversely affected. We note that this machine learning-inspired approach leads to good-quality spectra. Still, we also tested a more straightforward approach, based on a least-squares solution of the equation $\mathbf{s} = \mathbf{F} \mathbf{u}$ or, equivalently, setting $\mathbf{G}$ to be the Moore-Penrose pseudoinverse of $\mathbf{F}$. This approach gives inferior results with substantial spurious oscillations in the spectrum. This is discussed in the Supplementary Information.

\section{Conclusions}
\label{sec-conclusions}

In this work, we constructed a digital twin for a chiral sensing platform that uses helicity-preserving optical cavities to enhance light-matter interaction. In this approach, the price paid for the CD signal enhancement is the modification of the measured spectrum. The digital twin replicates each step of the actual experiment by performing a computer simulation and can thus be used to map different samples to measurable CD spectra, removing the uncertainty brought about by the modified spectrum. We then went further and showed that, using similar principles, it is possible to create a spectrometer based on helicity-preserving cavities. This spectrometer benefits from the signal enhancement by the cavities and can reconstruct the spectrum of the molecule being measured by computational means. Thus, by using the results of the digital twin, the drastic modification of the CD spectrum of the molecule is no obstacle to the unambiguous retrieval of the spectrum from the measurements.

Modern advances in computer hardware, algorithms, and simulation methods make it increasingly possible to use simulations as an integral part of the experimental process. Standard experiments, such as measuring a CD spectrum of molecules in solution, are normally straightforward because the setup is simple, and the results can be obtained from raw measurement data by analytical formulae. This is no longer the case with nanophotonic measurement devices, but the extra complexity can be managed with digital twins. In the case of a standard measurement with a known system that has to be repeated for many samples, the simulations could be performed in the background by intelligent computer software, leaving a straightforward interface for the user that requires no simulation expertise. Such systems appear as a logical next step in developing nanophotonics-based measurement technology.

\medskip
\textbf{Acknowledgements} \par
M.N. and C.R. acknowledge support by the KIT through the “Virtual Materials Design” (VIRTMAT) project. M.K. and C.R. acknowledge support by the Deutsche Forschungsgemeinschaft (DFG, German Research Foundation) under Germany’s Excellence Strategy via the Excellence Cluster 3D Matter Made to Order (EXC-2082/1-390761711) and from the Carl Zeiss Foundation via the CZF-Focus@HEiKA Program. M.K., C. H., and C.R. acknowledge funding by the Volkswagen Foundation. I.F.C. and C.R. acknowledge support by the Helmholtz Association via the Helmholtz program “Materials Systems Engineering” (MSE). M.N., M.K., and C.R. acknowledge support by the state of Baden-Württemberg through bwHPC and the German Research Foundation (DFG) through grant no. INST 40/575-1 FUGG (JUSTUS 2 cluster) and the HoreKa supercomputer funded by the Ministry of Science, Research and the Arts Baden-Württemberg and by the Federal Ministry of Education and Research. P.S. thanks the Hector Fellow Academy for their support. The authors are grateful to the company JCMwave for their free provision of the FEM Maxwell solver JCMsuite.

\medskip

%\bibliography{ChiralEnhancement.bib}

\end{document}

\title{Supplementary Information: A digital twin for a chiral sensing platform}
\date{\today}

\author{Markus Nyman*}
\affiliation{Institute of Nanotechnology, Karlsruhe Institute of Technology}
\email{markus.nyman@kit.edu, carsten.rockstuhl@kit.edu}

\author{Xavier Garcia-Santiago}
\affiliation{Institute of Nanotechnology, Karlsruhe Institute of Technology}
\author{Marjan Krsti\'c}
\affiliation{Institute of Theoretical Solid-State Physics, Karlsruhe Institute of Technology}

\author{Lukas Materne}
\affiliation{Institute of Theoretical Solid-State Physics, Karlsruhe Institute of Technology}

\author{Ivan Fernandez-Corbaton}
\affiliation{Institute of Nanotechnology, Karlsruhe Institute of Technology}

\author{Christof Holzer}
\affiliation{Institute of Theoretical Solid-State Physics, Karlsruhe Institute of Technology}

\author{Philip Scott}
\affiliation{Institute of Applied Physics, Karlsruhe Institute of Technology}

\author{Martin Wegener}
\affiliation{Institute of Nanotechnology, Karlsruhe Institute of Technology}
\affiliation{Institute of Applied Physics, Karlsruhe Institute of Technology}

\author{Willem Klopper}
\affiliation{Institute of Physical Chemistry, Karlsruhe Institute of Technology}

\author{Carsten Rockstuhl*}
\affiliation{Institute of Nanotechnology, Karlsruhe Institute of Technology}
\affiliation{Institute of Theoretical Solid-State Physics, Karlsruhe Institute of Technology}

\maketitle

\section{Quantum chemistry simulations}
The 1,1’-Bi-2-naphthol (BINOL) molecule was selected as a system to study cavity enhancement of CD signal in the infra-red part of the spectrum. Precise quantum chemistry simulations have been performed using density functional theory (DFT) for the ground state properties and its time-dependent variant (TD-DFT) for the excited state calculations as implemented into the development version of TURBOMOLE 7.7 \cite{TURBOMOLE2022} electronic structure calculation program. The workflow involved investigating the ground state structure of the BINOL molecule in the solvent surrounding, followed by calculations of dynamic polarizabilities of vibrational transitions. The molecular geometry of a single molecule was optimized using BFGS gradient minimization algorithm. The combination of hybrid PBE0 exchange-correlation functional \cite{adamoReliableDensityFunctional1999, ernzerhofAssessmentPerdewBurke1999}, Ahlrichs' def2-TZVP basis set \cite{weigendBalancedBasisSets2005, weigendAccurateCoulombfittingBasis2006} of a triple quality and Grimme's D4 dispersion correction \cite{caldeweyherExtensionD3Dispersion2017, caldeweyherGenerallyApplicableAtomiccharge2019} was employed to optimize the geometry and find total energy of the molecule. To accelerate the calculations, a resolution-of-identity (RI) method \cite{eichkornAuxiliaryBasisSets1995, eichkornAuxiliaryBasisSets1997} as well as multipole-accelerated resolution-of-identity (marij) \cite{sierkaFastEvaluationCoulomb2003} algorithms have been applied. The solvent effects on the geometry and properties of the BINOL molecule were included through an implicit Conductor-like screening model (COSMO) \cite{klamtCOSMONewApproach1993} for Chloroform. The refractive index n = 1.4441 and the dielectric constant $\varepsilon$ = 4.771 were set. The minimum on the ground state potential energy surface was confirmed by the analysis of the vibrational frequencies. Subsequently, damped complex dynamic polarizability tensors were calculated for vibrational and electronic transitions in the wavenumber window from 900-1900 cm\textsuperscript{-1} with a step of 1 cm\textsuperscript{-1}. Empirical damping at full-width at half maximum (FWHM) was set to 10 cm\textsuperscript{-1} for the Lorentzian-line shape to account for the thermal broadening effects to the discrete vibrational transitions. For the resulting spectra, the frequency in wavenumbers was scaled by a factor of 0.962 after the evaluation of the corresponding tensor components of the dynamic polarizability tensors 
$\alpha(\omega)$. 
Finally, calculated polarizability tensors of the molecule were used to construct T-matrices, later included into the digital twin of the entire optical device for the CD enhancement. 

\section{Beam-resonator interaction at oblique incidence}

Let us represent a normal-incidence optical beam as a collection of plane waves that have wave vectors \mbox{$\mathbf{k} = (k_x, k_y, \sqrt{(k^2 - k_x^2 - k_y^2})$}, where $k_x$ and $k_y$ can take values between $-k \sin(\theta)$ and $k \sin(\theta)$. Here, $\theta$ is the angular width of the beam, as in the main text, and the planar resonator through which the beam passes lies in the $xy$-plane. To represent an oblique-incidence beam, let us rotate these vectors about the $x$-axis by an angle $\theta_\text{inc}$. The resulting waves have
\begin{equation}
k_z' = k_y \sin(\theta_\text{inc}) + \sqrt{1-k_x^2-k_y^2}\cos(\theta_\text{inc}) \approx k_y \sin(\theta_\text{inc}) + \cos(\theta_\text{inc}).
\end{equation}
Here, the approximation $\sqrt{1-k_x^2-k_y^2} \approx 1$ was used because the maximum value of $k_x$ and $k_y$ is small if $\theta$ is small. The incidence angle of each plane wave with respect to the planar resonator is
\begin{equation}
\theta_\text{wave}(k_x,k_y) = \arccos(k_z'/k) \approx \theta_\text{inc} - k_y/k .
\end{equation}
Here, we approximated that $\theta << \theta_\text{inc}$. In this approximation, $\theta_\text{wave}$ does not depend on $k_x$. In the main text, we use $\theta = $~0.02~rad, and the beams inside the resonator have $\theta_\text{inc} =$ 80$^\circ$, making these approximations good.

Now, assume that the resonator is on resonance at the incidence angle $\theta_\text{inc}$ and that its transmittance has, as a function of the wave incidence angle, a full width at half maximum of $\Delta \theta$. Let us assume that all plane waves with $|\theta_\text{wave}(k_x,k_y) - \theta_\text{inc}| < \Delta \theta / 2$ are admitted by the resonator and the rest are reflected with little interaction. Clearly, this condition can be written as $|k_y/k| < \Delta \theta / 2$, and does not depend on $k_x$ at all. Meanwhile, the beam contains plane waves with $|k_y/k| < \theta$. Therefore, the fraction of the beam's plane waves satisfying this condition is $\Delta \theta / (2 \theta)$ if $\theta > \Delta \theta / 2$. If each plane wave has the same intensity, then the fraction of beam energy admitted into the resonator is inversely proportional to $\theta$. Although this argument is made for the transmission of beams through the resonator, it holds for the field enhancement as well.

This explains the behavior observed in Fig.~4(a) of the main text. In the helicity-preserving cavity, it is the diffraction by the cylinder array that creates beams that propagate at oblique angles of incidence.

\section{Spectral reconstruction}

To construct the matrix $\mathbf{G}$ that reconstructs a spectrum $\mathbf{u}$ out of measured cavity signals $\mathbf{s}$ using $\mathbf{u} = \mathbf{G} \mathbf{s}$, we regard $\mathbf{G}$ as a single-layer neural network with a linear activation function. The coefficients of $\mathbf{G}$ are trained using the TensorFlow 2.3.4 library. The data set is 10.000 spectra, each of which is a sum of 5 Lorentzian functions with a fixed spectral width of \SI{0.015}{\micro\meter} and a randomly chosen peak wavelength between \SI{6.0}{\micro\meter} and \SI{7.6}{\micro\meter}. Out of these, 6.000 spectra are used as training data, 2.000 as validation data, and 2.000 as test data. The training process uses the Adam algorithm, minimizing the mean absolute error between the real and reconstructed spectra. The training is performed for 200 epochs, in batches of 32 spectra, the learning rate is set to $10^{-6}$ and an L1 regularizer is used (with regularization parameter $10^{-6}$). Importantly, the optimization is started from $\mathbf{G} = \mathbf{0}$ instead of randomly chosen coefficients. This avoids the addition of noise.

\begin{figure}
\centering
\includegraphics[width=0.6\textwidth]{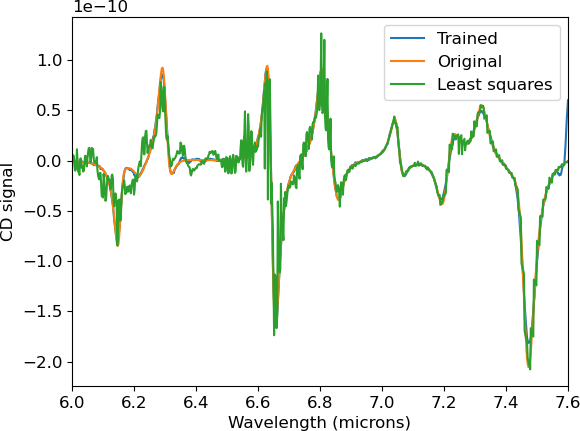}
\caption{The original spectrum of BINOL (orange) compared to the reconstructed spectrum using the machine learning-inspired approach (blue) and the reconstructed spectrum obtained from a least-squares solution (green).}
\label{fig-supp-leastsquares}
\end{figure}

In addition to the machine learning-inspired approach detailed here, we performed the reconstruction using the least-squares solution to $\mathbf{s} = \mathbf{F} \mathbf{u}$, utilizing both the \texttt{scipy.linalg.lstsq} function of the SciPy library or letting $\mathbf{G}$ be the Moore-Penrose pseudoinverse of $\mathbf{F}$ (both approaches yield the same result). In these simulations we also used for each cavity a \SI{0.1}{\micro\meter} wide bandpass filter centered around that cavity's central frequency. The filters reject most of the side peaks which should improve the reconstruction for all algorithms. Figure~\ref{fig-supp-leastsquares} compares the original spectrum of BINOL and the reconstructed spectra obtained from both the machine learning-inspired approach and the least squares solution. We see that while the former reproduces the original spectrum very well, the latter introduces copious amounts of high-frequency spurious oscillations. This may be because the training data for the machine learning approach only contains sums of Lorentzians with a constant broadening parameter and not completely arbitrary functions. This would ensure that, e.g., sharp spikes and fast oscillations would not be present in the reconstructed spectra.

% We have 'function of spacing' results which do not yield perfect reconstruction if there are not enough cavities. But I don't think they need to be here. If a reviewer asks for it we have it.

\newpage

\section{Cartesian coordinates of the optimized BINOL molecule}
36
\\
\\
C     3.7023012    0.2185229    2.1956654\\
C     3.8240576   -0.7763305    1.2644034\\
C     2.8101792   -1.0023606    0.3086282\\
C     1.6453292   -0.1858818    0.3161842\\
C     1.5532639    0.8374623    1.2878914\\
C     2.5536195    1.0327674    2.2016349\\
C     2.9210930   -2.0203829   -0.6614660\\
C     1.9298814   -2.2215543   -1.5766082\\
C     0.7734493   -1.4148334   -1.5634115\\
C     0.6163042   -0.4101011   -0.6351720\\
C    -0.6124936    0.4139542   -0.6379618\\
C    -1.6432529    0.1900828    0.3116113\\
C    -2.8130125    0.9993029    0.2946028\\
C    -2.9264720    2.0105583   -0.6822773\\
C    -1.9326243    2.2126177   -1.5943872\\
C    -0.7715895    1.4126637   -1.5722317\\
C    -3.8293061    0.7725061    1.2476198\\
C    -3.7049152   -0.2157462    2.1855544\\
C    -2.5510576   -1.0224510    2.2012652\\
C    -1.5485550   -0.8266385    1.2900267\\
O     0.2192441    1.6194806   -2.4705167\\
O    -0.2154918   -1.6218034   -2.4637713\\
H    -3.8172417    2.6291935   -0.7004546\\
H     3.8077887   -2.6449600   -0.6718380\\
H    -2.0222832    2.9920116   -2.3452646\\
H     2.0175953   -3.0060516   -2.3223620\\
H     0.6742033    1.4702946    1.2995688\\
H    -0.6655701   -1.4537748    1.3094351\\
H     4.7054502   -1.4092715    1.2453185\\
H    -4.7147709    1.3994260    1.2207887\\
H     2.4619132    1.8233972    2.9378892\\
H    -2.4569663   -1.8076833    2.9430057\\
H    -4.4902967   -0.3817921    2.9137609\\
H     4.4857629    0.3839298    2.9260971\\
H    -0.0232724    2.3468504   -3.0526374\\
H     0.0239055   -2.3557347   -3.0389112\\

\section{TURBOMOLE control file for the dynamic polarizabilities}

{\$}cosmo \\
  epsilon=    4.711 \\
  rsolv= 1.30 \\
  refind=   1.4441 \\
{\$}cosmo\_atoms \\
\# radii in Angstrom units \\
c  1-20                                                                        \ \\
   radius=  2.0000 \\
o  21-22                                                                       \ \\
   radius=  1.7200 \\
h  23-36                                                                       \ \\
   radius=  1.3000 \\
{\$}cosmo\_out file=out.ccf \\
{\$}cosmo\_data file=cosmo\_transfer.tmp \\
{\$}title \\
{\$}symmetry c1 \\
{\$}coord    file=coord \\
{\$}optimize \\
 internal   off \\
 redundant  off \\
 cartesian  on \\
{\$}atoms \\
    basis =def2-TZVP \\
    jbas  =def2-TZVP \\
{\$}basis    file=basis \\
{\$}scfmo   file=mos \\
{\$}scfiterlimit     1000 \\
{\$}scfdamp   start=0.300  step=0.050  min=0.100 \\
{\$}scfdump \\
{\$}scfdiis \\
{\$}maxcor    500 MiB  per\_core \\
{\$}scforbitalshift automatic=.1 \\
{\$}energy    file=energy \\
{\$}grad    file=gradient \\
{\$}dft \\
   functional   pbe0 \\ % was no grid specified? -> m3 would be the default than
{\$}scfconv   7 \\
{\$}ricore     2048 \\
{\$}rij \\
{\$}jbas    file=auxbasis \\
{\$}marij \\
{\$}disp4 \\
{\$}mgiao \\
{\$}vcd \\
{\$}vibcd dynpol 1/cm \\
 900, 1900, 1001 \\
{\$}damped\_response 5 1/cm \\ % new convention: seems that 5 has been used (= Lukas 10)
{\$}magnetic\_response \\
{\$}rundimensions \\
   natoms=36 \\
{\$}closed shells \\
 a       1-75                                   ( 2 ) \\
{\$}last step     define \\
{\$}end \\

%\bibliography{ChiralEnhancement.bib}
%\printbibliography